# Proximity-Induced Superconductivity in a Topological Crystalline Insulator

Bryan Rachmilowitz[1,*], He Zhao[1,*], Hong Li[1], Alex LaFleur[1], J. Schneeloch[2], Ruidan Zhong[2], Genda Gu[2] and Ilija Zeljkovic[1,¶]

[1]Department of Physics, Boston College, 140 Commonwealth Ave, Chestnut Hill, MA 02467;  [2]Brookhaven National Laboratory, Upton, NY 11973; [¶]Corresponding author: ilija.zeljkovic@bc.edu; [*]Equal contribution.

**Abstract.**

Superconducting topological crystalline insulators (TCI) are predicted to host new topological phases protected by crystalline symmetries, but available materials are insufficiently suitable for surface studies. To induce superconductivity at the surface of a prototypical TCI SnTe, we use molecular beam epitaxy to grow a heterostructure of SnTe and a high-$T_c$ superconductor Fe(Te,Se), utilizing a "buffer" layer to bridge the large lattice mismatch between SnTe and Fe(Te,Se). Using low-temperature scanning tunneling microscopy and spectroscopy, we measure a prominent spectral gap on the surface of SnTe, and demonstrate its superconducting origin by its dependence on temperature and magnetic field. Our work provides a new platform for atomic-scale investigations of emergent topological phenomena in superconducting TCIs.

**Introduction.**

Topological crystalline insulators (TCIs) are a subclass of topological materials in which the emergence of non-trivial surface states is intimately tied to a discrete set of crystalline symmetries [1]. Numerous types of symmetries can in principle lead to a TCI phase [1], but only one class of TCIs based on the reflection symmetry with respect to the (110) mirror plane [2] has so far been experimentally realized in rock-salt (Pb,Sn)Se and (Pb,Sn)Te [3–6]. The electronic band structure of these TCIs consists of multiple Dirac fermions, tunable by temperature [3,6], chemical composition [7–9] and various types of strain [9–12], and provides a rich playground for uncovering new physics [13–17]. For example, a single unit cell thick film of SnTe is shown to exhibit room-temperature ferroelectricity [14], while a bulk single crystal of (Pb,Sn)Se is reported to host 1D edge modes propagating along step edges [15,17]. Theory predicts that if a TCI undergoes a superconducting transition, a new topological superconducting phase could emerge [18–21] distinct from that in proximitized $Z_2$ topological insulators Bi$_2$(Se,Te)$_3$ [22–28]. The difference is rooted in the unique mirror symmetry protection and multiple Dirac fermions present at the surface of TCIs, which are expected to facilitate novel phenomena emerging at the surface. These include multiple symmetry-protected Majorana zero modes inside a single vortex core [20] and tunable Andreev bound states [29]. The bottleneck in exploring these lies in the synthesis of high-quality surfaces of superconducting TCIs.

In analogy to achieving superconductivity in $Z_2$ topological insulators [30], alloying has been successfully used to induce bulk superconductivity in TCIs [31]. Bulk single crystals of Sn$_{1-x}$In$_x$Te exhibit superconductivity up to ~4.5 K [31] and maintain the topological nature of the Dirac surface states [32–34]. However, the difficulty in obtaining a large, flat surface of Sn$_{1-x}$In$_x$Te by cleaving has hindered nanoscale explorations of the superconducting TCI phase. An alternative method to induce superconductivity in a material, without introducing chemical disorder, could entail using the proximity effect [35]. This approach typically involves deposition or exfoliation of thin films on top of superconducting substrates, and it has been widely applied to $Z_2$ topological insulators [22–28]. The initial efforts to create TCI/superconductor heterostructures appear promising [36], but achieving atomically flat interfaces in this geometry has been extremely difficult. This is in large part due to a strong intra-layer bonding of (Pb,Sn)Se and (Pb,Sn)Te, which prevents their mechanical exfoliation and severely limits the choice of viable substrates for epitaxial thin film growth. Unlike van der Waals topological insulators Bi$_2$(Se,Te)$_3$ that can be grown on a slew of substrates nearly irrespective of the structural compatibility [22,37], TCI thin films are known to be strongly susceptible to warping [11,12,38]. Thus, the growth of atomically flat TCIs is best achieved on the substrates with a closely matched in-plane lattice structure [39,40], whereas their growth on commonly used superconducting substrates, such as NbSe$_2$ and elemental superconductors where the lattice mismatch is inevitably much larger than one percent, remains challenging. In this work, by the use of ultrathin "buffer" layers to bridge the structural mismatch, we successfully grow atomically flat SnTe thin films on top of a high-$T_c$ superconductor Fe(Te,Se), and find the proximity-induced superconductivity at the surface of SnTe.

**Results.**

*MBE growth of heterostructures.*

The starting point for our heterostructure growth is a bulk single crystal of superconducting FeTe$_{0.55}$Se$_{0.45}$ (Fe(Te,Se)) with $T_c$ ~ 14K, cleaved in ultra-high vacuum to expose a clean surface. Scanning tunneling microscopy (STM) topographs of the Fe(Te,Se) surface show a square lattice with $a_0$ ~ 3.9 Å (Fig. 1b), which demonstrates that the Fe(Te,Se) crystals cleave along the (001) direction. Instead of growing SnTe thin films directly on top of Fe(Te,Se), we first deposit one quintuple layer (QL) of Bi$_2$Te$_3$ using molecular beam epitaxy (MBE) (Fig. 1a). This crucial layer serves to bridge the structural difference between Fe(Te,Se) and SnTe. Despite the obvious incompatibility between the in-plane lattices of Fe(Te,Se) (square lattice with $a_0$~3.9 Å) and Bi$_2$Te$_3$ (hexagonal lattice with $a_0$~4.4 Å), weak van der Waals bonding at the interface allows for the epitaxial growth of Bi$_2$Te$_3$ [27]. At the same time, Bi$_2$Te$_3$ can serve as nearly an ideal substrate for the growth of SnTe along the (111) direction, because the difference between the in-plane lattice constants of the two materials is only ~1.7% [40], comparable to that between SnTe and the commonly used BaF$_2$ substrates [34]. This allows us to deposit SnTe thin films of varying thicknesses on top of 1 QL Bi$_2$Te$_3$ to successfully complete the heterostructure shown in Fig. 1a. We choose the 1 QL thickness of the Bi$_2$Te$_3$ buffer layer (~1 nm thick) to minimize the separation between SnTe and the superconducting substrate, which should in principle maximize the proximity-induced pairing correlations at the surface of SnTe [41].

*Nanoscale structural and electronic characterization.*

STM topographs of SnTe(111) surface show a hexagonal lattice of Te atoms (Fig. 1d), qualitatively similar to those reported by previous experiments [40]. The topographs are also bias-dependent, appearing more inhomogeneous at higher bias (Fig. S1 [49]). Importantly, they are clearly distinct from the characteristic topographs of underlying 1QL Bi$_2$Te$_3$ (Fig. 1c). We further confirm the nature of the terminating layer of our heterostructure by measuring the step height in the STM topographs to be ~0.4 nm (Fig. 1e), which equals the height of one SnTe(111) bilayer (BL).

We proceed to use low-temperature scanning tunneling spectroscopy to characterize the electronic properties of the film. We present data on three different SnTe films with nominal thicknesses of ~3 BL, ~6 BL and ~14 BL, labeled A, B and C, respectively. From differential conductance (*dI/dV*) spectra acquired over a large energy range, we can estimate the top of the valence band (VBT), which is seen as a sharp upturn in conductance at negative energies (Fig. 2a). With increased SnTe film thickness, Fermi level shifts down towards the VBT, and it falls just above the VBT in the thickest film. Finite *dI/dV* conductance above the VBT would be consistent with the existence of surface states spanning the Fermi level.

Quasiparticle interference (QPI) measurements provide further evidence supporting this scenario (Fig. 3), as the observed QPI morphology is qualitatively consistent with the expected Dirac cone structure of the SnTe(111) surface state [42]. The surface state of bulk SnTe(111) consists of a Dirac cone at Γ and another one at each M point (Fig. 3(a)), with the cones at the two different *k*-space position possibly slightly offset in energy [42]. Given this surface state structure, there are three dominant inter-band scattering wave vectors, which are schematically depicted in Fig. 3(a). In the Fourier transforms of STM dI/dV maps in our samples, we observe diffuse signatures at the positions corresponding to all three of these scattering channels, at several different energies across the Fermi level (Fig. 3(b-d)). We note that $k_z$-dispersing bulk bands are typically not observed in QPI measurements [43]. The emergence of QPI vectors at these positions is consistent with the existence of the Dirac surface states in our samples. We also note that in the 2D limit, hybridization of the top and the bottom surface state can in principle lead to a gap opening at the Dirac point, but the topological nature of these states is expected to remain intact in a large range of thicknesses [44].

Next, we look for signatures of induced superconductivity by measuring *dI/dV* spectra over a narrow energy range near the Fermi level. Before depositing SnTe, average *dI/dV* spectrum on top of 1QL $Bi_2Te_3$/Fe(Te,Se) shows a clear gap in the density of states symmetric with respect to the Fermi level, consistent with a proximity-induced superconducting gap at the surface of $Bi_2Te_3$ thoroughly explored in our previous work [27]. *dI/dV* spectra acquired on the surface of SnTe(111) in samples A and B also show a prominent, symmetric spectral gap in the density of states (Fig. 2b,c). Spectral gap variation at the surface of the two samples, both a comparable distance away from the bulk superconductor Fe(Te,Se), is likely due to different STM tips used, as well as the variations in Se:Te ratio and the concentration of excess interstitial Fe across different Fe(Te,Se) substrates[45,46,47]. Nevertheless, within a single heterostructure, the measured gap is spatially homogeneous over the nanoscale region measured, with little variation in the spectral shape (Fig. 2e). A much thicker sample C only shows a small suppression in the density-of-states near the Fermi level (Fig. 2d). This would be an expected trend for a proximity-induced superconducting gap, where superconducting correlations decay away from the bulk superconductor [41]. The quick suppression of the gap away from the interface is possibly due to the relatively short-coherence length of Fe(Te,Se)[28,47].

To further investigate the gap observed on the surface of SnTe(111), we focus on a 40 nm square region of sample A, and track the average *dI/dV* spectrum as a function of temperature (Fig. 2f). The measured gap becomes shallower with increased temperature, and ultimately disappears at ~ 12 K. This is comparable to the bulk $T_c$ ~ 14 K of the Fe(Te,Se) substrate, and provides additional support for the superconducting origin of the gap. We fit the gap magnitude using a thermally-broadened BCS function (Fig. S2 [49]), which provides a good fit to the experimental data [49]. We find that the extracted gap magnitude as a function of temperature closely follows the BCS trend (Fig. 2g).

Lastly, we use spectroscopic imaging STM to spatially map the differential conductance at varying magnetic fields applied perpendicular to the sample surface. In type-II superconductors such as Fe(Te,Se), magnetic field will penetrate the material in quantized vortices. If the surface of SnTe is indeed superconducting, we would expect to observe Abrikosov vortices as localized regions of low *dI/dV* conductance in STM *dI/dV* maps acquired at energies of the superconducting gap [50]. Fig. 4a-c shows *dI/dV* maps acquired over the identical region of SnTe in varying magnetic field. All images show a clear vortex lattice, with the number of vortices scaling with the applied magnetic field, confirming the induced superconductivity at the surface of SnTe. Importantly, the induced gap is larger and persists to higher temperatures compared to what has been achieved in proximity experiments in TCIs so far using low-temperature s-wave superconductors [36]. Interestingly, the vortex cores in superconducting SnTe exhibit not only the expected suppression of the coherence peaks in *dI/dV* spectra, but also an intriguing peak in *dI/dV* conductance centered at zero energy (Fig. 4d). Similar zero-bias conductance peaks have now been reported in vortices of several related systems, including a fraction of vortices measured on superconducting Fe(Te,Se)

[51], Fe impurities on the surface of Fe(Te,Se) [52] and a monolayer of FeSe [53], and $Bi_2Te_3$/Fe(Te,Se) heterostructures [27,28] (Fig. S3 [49]).

**Discussion.**

We have demonstrated that $Bi_2Te_3$ buffer layer enables the realization of a novel heterostructure involving a topological crystalline insulator SnTe and a high-$T_c$ superconductor Fe(Te,Se). Temperature and magnetic field dependent scanning tunneling microscopy/spectroscopy measurements of the SnTe surface demonstrate induced superconductivity in SnTe, with the highest onset temperature and the largest energy gap in superconducting TCIs to-date. The use of the same buffer layer can easily be extended to couple TCIs to other unconventional superconductors or magnetic materials. Future experiments directly demonstrating spin-momentum locking and the superconducting gap in the surface states may shed light on the existence of topological superconductivity in proximitized TCIs. Moreover, spin-polarized STM [26] could be used to explore the potential emergence of Majorana zero modes in vortices. Overall, our experiments provide a new platform to study the interplay of crystalline symmetries, topology and superconductivity in a single material.

**Methods.**

$FeTe_{1-x}Se_x$ (x~0.45) bulk single crystals were grown using the self-flux method. During the growth process, RHEED pattern (obtained using a 15 keV RHEED gun by Sentys Inc) was continuously monitored to establish the morphology of the surface. $Bi_2Te_3$ film was grown by co-evaporating 99.999% pure Bismuth and 99.99% pure Tellurium from K-cells (Sentys Inc) in Bi:Te flux ratio of ~ 1:10. SnTe film was grown by co-evaporating 99.999% Tin and 99.99% Tellurium in Sn:Te~ 1:10 flux ratio. $Bi_2Te_3$ (SnTe) films were grown at the rate of 4 (2) minutes per nominally calculated QL (bilayer). Sample A was grown at ~200 °C ($Bi_2Te_3$) and ~300 °C (SnTe); sample B was grown at ~250 °C ($Bi_2Te_3$) and ~300 °C (SnTe), but post-annealed to at least 320 °C which evaporated the $Bi_2Te_3$ layer off (Fig. S4 [49]); sample C was grown at ~180 °C (both $Bi_2Te_3$ and SnTe), and post annealed for 30 minutes in vacuum at ~270 °C. Typical post-growth RHEED pattern of our heterostructures exhibits a streaky pattern characteristic of the layer-by layer MBE growth (Fig. S5 [49]). After the growth was completed, the heterostructure was transferred from the MBE to the STM within one hour, using a vacuum suitcase chamber held at ~$10^{-11}$ Torr base pressure, which can be directly connected to either MBE or STM chambers. Therefore, we emphasize that our material is only exposed to UHV conditions during the entire process from the start of the MBE growth to the completion of STM measurement.

STM measurements were acquired using Unisoku USM1300 STM at the base temperature of ~4.5 K (with the exception of temperature dependent data in Fig. 2g). STM tips used were home-made chemically etched metallic Cr tips. All spectroscopic measurements have been taken using a standard lock-in technique at 915 Hz frequency and a varying bias excitation as detailed in the figure captions.

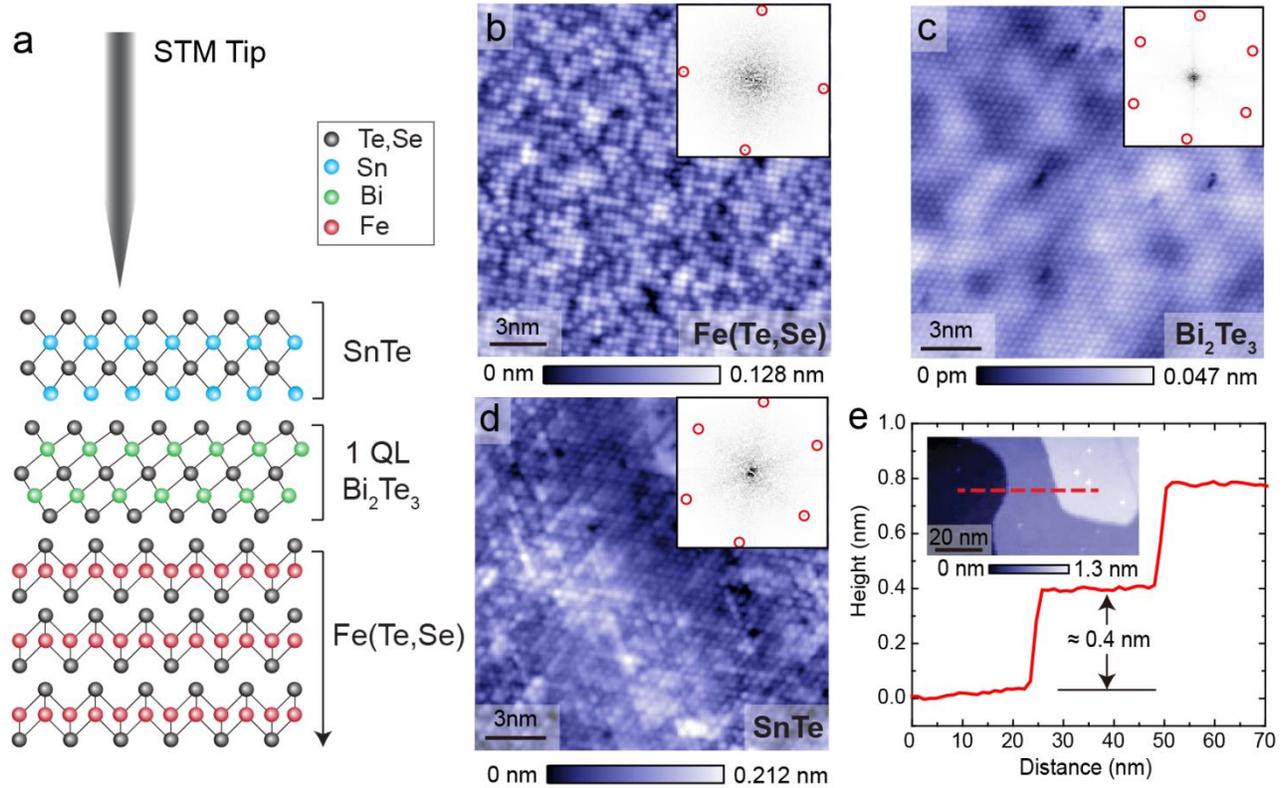

**Figure 1.** (a) Schematic of SnTe/Bi$_2$Te$_3$/Fe(Te,Se) heterostructure. STM topograph showing (b) exposed Fe(Te,Se) substrate, (c) 1 QL Bi$_2$Te$_3$ buffer layer, and the (d) topmost SnTe layer. Insets in (b-d) are the respective Fourier transforms showing their lattice symmetries. (e) Height profile taken along the red dashed line in the embedded STM topograph. The step height is consistent with consecutive bilayers of SnTe. STM setup conditions: (b) $I_{set}$ = 15 pA, $V_{sample}$ = 10 mV; (c) $I_{set}$ = 30 pA, $V_{sample}$ = 6 mV; (d) $I_{set}$ = 40 pA, $V_{sample}$ = 40 mV; (e) $I_{set}$ = 10 pA, $V_{sample}$ = -500 mV.

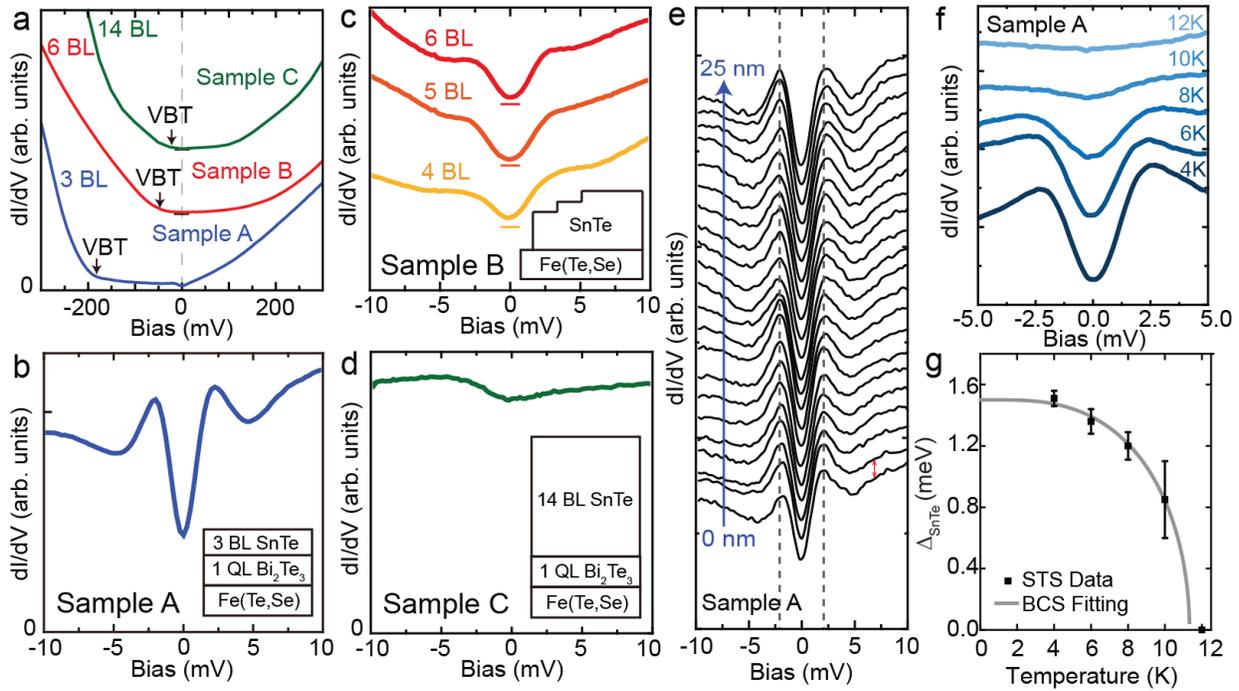

**Figure 2.** (a) Long range average *dI/dV* spectra on sample A (blue), sample B (red) and sample C (green). VBT denotes on the approximate position of the top of the valence band. (b-d) Average *dI/dV* spectra on the surface of (b) sample A, (c) sample B and (d) sample C, with a schematic of their respective heterostructures (inset in in the lower right corner). (e) A series of *dI/dV* spectra, offset for clarity, along a 25 nm line taken on the surface of sample A. (f) Average *dI/dV* spectra, offset for clarity, as a function of temperature, showing the gap closing at ~12 K. (g) Spectral gap $\Delta_{SnTe}$ as a function of temperature extracted from spectra in (f) (black squares), and the overlaid BCS trend (grey line). STM setup condition: (a) Sample A: $I_{set}$ = 100 pA, $V_{sample}$ = 300 mV, $V_{exc}$ = 10 mV (zero-to-peak); Sample B: $I_{set}$ = 200 pA, $V_{sample}$ = 400 mV, $V_{exc}$ = 10 mV; Sample C: $I_{set}$ = 100 pA, $V_{sample}$ = 300 mV, $V_{exc}$ = 10 mV; (b) $I_{set}$ = 40 pA, $V_{sample}$ = 10 mV, $V_{exc}$ = 0.2 mV; (c) 4 BL: $I_{set}$ = 100 pA, $V_{sample}$ = 10 mV, $V_{exc}$ = 0.2 mV; 5 BL: $I_{set}$ = 60 pA, $V_{sample}$ = 10 mV, $V_{exc}$ = 0.2 mV; 6 BL: $I_{set}$ = 60 pA, $V_{sample}$ = 10 mV, $V_{exc}$ = 0.2 mV; (d) $I_{set}$ = 150 pA, $V_{sample}$ = 10 mV, $V_{exc}$ = 0.3 mV; (e) $I_{set}$ = 40 pA, $V_{sample}$ = 10 mV, $V_{exc}$ = 0.2 mV, (f) $I_{set}$ = 30 pA, $V_{sample}$ = 5 mV, $V_{exc}$ = 0.2 mV.

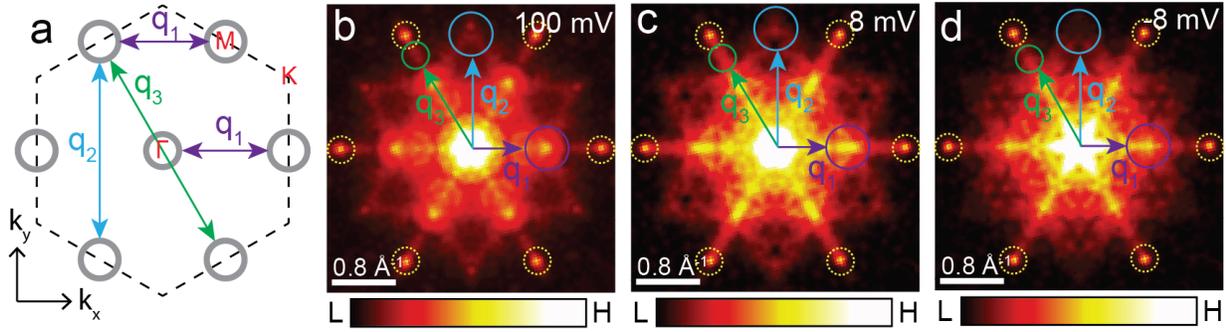

**Figure 3.** Quasiparticle interference (QPI) imaging of SnTe(111) surface of sample C. (a) The schematic of the SnTe (111) constant energy contour, where $q_1$ (purple), $q_2$ (blue) and $q_3$ (green) denote dominant scattering vectors. The dashed hexagon in (a) denotes the 1st Brillouin zone. (b-d) Fourier transforms of dI/dV maps acquired at 100 mV, 8 mV and -8 mV bias, respectively. The peaks in the Fourier transform in (b) circled in purple, blue and green correspond to the scattering channels denoted in panel (a). Yellow dashed circles in (b-d) denote the atomic Bragg peaks associated with the Te lattice. STM setup conditions: (b) $I_{set}$ = 200 pA, $V_{sample}$ = 100 mV, $V_{exc}$ = 2 mV; (c) $I_{set}$ = 200 pA, $V_{sample}$ = -8 mV, $V_{exc}$ = 2 mV; (d) $I_{set}$ = 200 pA, $V_{sample}$ = 8 mV, $V_{exc}$ = 2 mV.

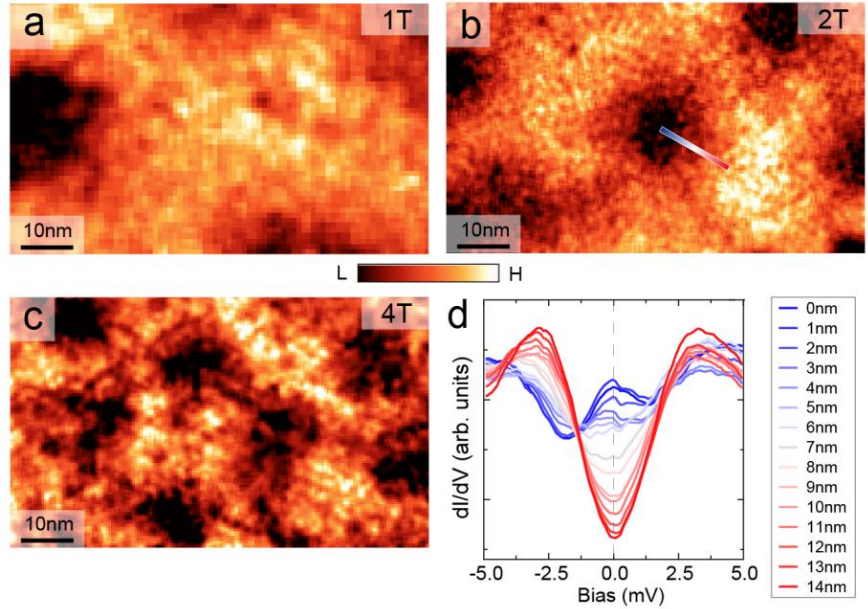

**Figure 4.** (a-c) *dI/dV* maps over the same region of the sample at 1 T, 2 T and 4 T magnetic fields, respectively. All fields are applied perpendicular to the surface of the sample. (d) Radially-averaged *dI/dV* spectra as a function of distance away from the vortex core in (b). Zero bias peak is visible in the vortex core (darkest blue). STM setup condition: (a) $I_{set}$ = 4 pA, $V_{sample}$ = 2 mV, $V_{exc}$ = 0.4 mV; (b) $I_{set}$ = 8 pA, $V_{sample}$ = 2 mV, $V_{exc}$ = 0.4 mV.

**Acknowledgements**

I.Z. gratefully acknowledges the support from Army Research Office Grant No. W911NF-17-1-0399 (MBE growth) and National Science Foundation Grant No. NSF-DMR-1654041 (STM characterization). The work in Brookhaven was supported by the Office of Science, US Department of Energy under Contract No. DE-SC0012704. J.S. and R.D.Z. were supported by the Center for Emergent Superconductivity, an Energy Frontier Research Center funded by the U.S. Department of Energy, Office of Science.